\documentclass[12pt,preprint]{aastex}
\usepackage{natbib}
\newcommand{\idn}{ads/sa.spitzer\#0007427328}

\slugcomment{}%

\begin{document}

\shorttitle{Excesses in M47}
\shortauthors{N. Gorlova et al.}

%________________________________________________________________
\title{New debris disk candidates:\\
24$\micron$ stellar excesses at 100 Myr}


\author{Nadya Gorlova$^1$, Deborah L. Padgett$^2$, George H. Rieke$^1$, James Muzerolle$^1$, Jane E. Morrison$^1$,
Karl D. Gordon$^1$, Chad W. Engelbracht$^1$, Dean C. Hines$^{1, 3}$, Joannah C. Hinz$^1$, Alberto Noriega-Crespo$^2$, Luisa Rebull$^2$, 
John A. Stansberry$^1$, Karl R. Stapelfeldt$^4$, Kate Y. L. Su$^1$, Erick T. Young$^1$ }

\affil{$^1$Steward Observatory, University of Arizona, Tucson, AZ 85721}
\email {ngorlova@as.arizona.edu}
\affil{$^2$Spitzer Science Center, California Institute of Technology, Pasadena,CA 91125}
\affil{$^3$Space Science Institute, Boulder, Colorado 80301}
\affil{$^4$Jet Propulsion Laboratory, California Institute of Technology, Pasadena, CA 91109.}



\begin{abstract}
 Sixty three members of the 100 Myr old open cluster M47 (NGC 2422) have been 
detected 
at 24$\micron$ with {\it Spitzer}. 
The Be star V 378 Pup shows an excess both in the near-infrared and at 24$\mu$m ($K-[24] = 
2.4$ mag),
probably due to free-free emission from the gaseous envelope. Seven other early-type
stars show smaller excesses, $K-[24] = 0.6-0.9$.

Among late-type stars, two show large excesses: P922 -- a K1V star with $K-[24] 
= 1.08 \pm 0.11$ and 
P1121 -- an F9V star with $K-[24] = 3.72 \pm 0.02$. 
P1121 is the first known main-sequence star showing an excess comparable to 
that of $\beta$ Pic, which may indicate 
the presence of an exceptionally massive debris disk. It is possible that a 
major
planetesimal collision has occurred in this system, consistent with the few
hundred Myr time scales
estimated for the clearing of the Solar System.
\end{abstract}


%% subject header.  Objects should be in the appropriate "individual"
%% headers (e.g. quasars: individual, stars: individual, etc.) with the
%% additional provision that the total number of headers, including each
%% individual object, not exceed six.  The \objectname{} macro, and its
%% alias \object{}, is used to mark each object.  The macro takes the object
%% name as its primary argument.  This name will appear in the paper
%% and serve as the link's anchor in the electronic edition if the name
%% is recognized by the data centers.  The macro also takes an optional
%% argument in parentheses in cases where the data center identification
%% differs from what is to be printed in the paper.


%% \keywords{globular clusters: general ---
%% globular clusters: individual(\objectname{NGC 6397},
%% \object{NGC 6624}, \objectname[M 15]{NGC 7078},
%% \object[Cl 1938-341]{Terzan 8})}


\section{Introduction}

  A knowledge of circumstellar disk evolution is crucial to 
understanding the planet formation process. The current paradigm
predicts that optically thick accretion disks
left over from star formation should dissipate as material is accreted, 
photoevaporated, or driven out of the system, leaving a remnant or debris disk formed by collisions 
between large bodies in a nascent planetary system \citep{Hollenbach2000,Wyatt2002}. 
Studies of how these circumstellar debris disks
evolve in the planet-forming zone from 1 to 10 AU, where the
disk emission emerges in the mid- and far-infrared, have been limited
by the sensitivity of the previous space-based infrared missions, IRAS and ISO.
The small number of F and A stars within
their reach has undermined attempts to
determine the frequency of excess emission. Claims of 
decreasing amounts of excess emission with
age \citep{Spangler2001} or an upper limit to the age of stars
with debris disks \citep{Habing2001} have been disputed by
others who question the S/N of the excesses in question and/or
the reliability of age estimates for nearby field stars \citep{Decin2003}.

The solution to this problem is to survey for debris disks 
around stars in young clusters of reliable and disparate ages.
The high sensitivity of the {\it Spitzer} MIPS instrument at last puts
this goal within reach for nearby young clusters. The use
of M47, a young Galactic open cluster, as a calibration target during the {\it 
Spitzer}
in-orbit checkout has provided an early opportunity to perform
such a survey.

   Messier 47 (NGC 2422) is a reasonably compact (core radius 30 arcmin, 
\citet{Barbera2002})
cluster of several hundred stars \citep{Ishmukhamedov1967,Mermilliod1986, 
Prisinzano2003} 
at galactic latitude 3$\degr $ and distance
around 450pc (see discussion in Sec \ref{memb}). Its age of
$80 \pm 20 $ Myr \citep*{RojoArellano1997, vanRensbergen1978}, 
is approximately the same as that of the Pleiades. Stromgren photometry of M47 indicates that
its metallicity is also similar to that of the Pleiades and $\alpha$ Per
clusters \citep{Nissen1988}. Despite its
proximity to the galactic plane, M47 has a small measured reddening of
$E(B-V) = 0.05 - 0.07$ \citep{Shobbrook1984, Nissen1988}.

\section{New Observations}

\subsection{Spitzer Measurements}

\begin{figure}  
\plotone{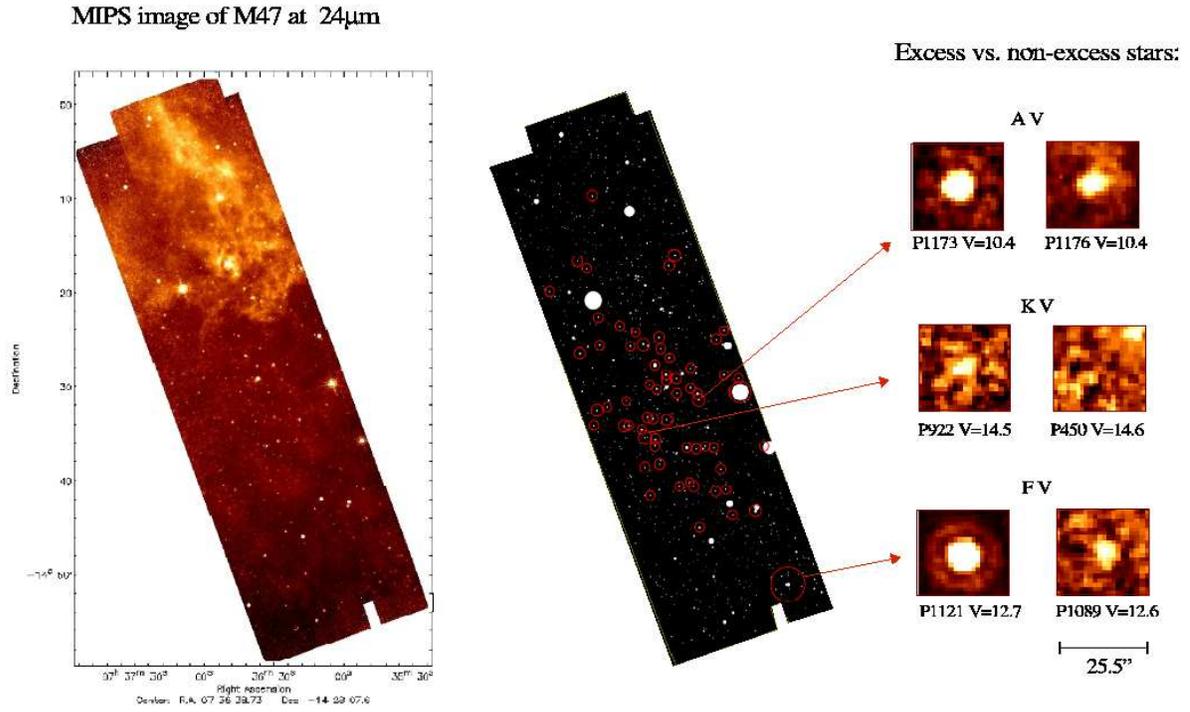}  
\caption{Left panel -- M47 mosaic. 
Central panel -- sources detected at 24$\mu$m,
red circles represent 24$\mu$m excess in M47 members (size $\sim K-[24]$). Right panel -- magnified images
of some excess stars (left column, arrows pointing to the position on the map),
compared to similar magnitude $V$, $B-V$ non-excess stars (right column).}\label{fig1}
\end{figure}

A region around M47 was observed with MIPS \citep{Rieke2004}
as a verification
of the scan map parameters during the in-orbit checkout period (\idn).
The sky coverage is non-uniform: the effective
exposure time varies over the combined $20' \times 60'$ map, from a minimum
of 100 seconds at the edges to a maximum of about 600 seconds 
in the central $\sim5' \times 60'$ strip.
The data were reduced using the instrument team Data Analysis
Tool \citep{Gordon2004}. 
The 24$\micron$ mosaic is shown in Fig. \ref{fig1}. It
is oversampled by a factor of two, i.e., the pixel size is 1.27$''$. The sky coverage was
not optimized for the 70$\micron$ array, and only a few (possibly extragalactic) 
sources were detected at this wavelength. Thus, all our analysis 
has focused on the 24$\micron$ data.

Aperture photometry was performed using the DAOPHOT package \citep{Stetson1987}.
The signal aperture radius and inner and outer sky radii 
were set at 3.5, 9 and 18 pixels (4.5$''$, 11.5$''$, and 23$''$), respectively. 
A 0.73 mag aperture correction was applied to the measured values after 
examining 
the trend of the signal with aperture size on well isolated stars and
in synthetic images. Final magnitudes were put on a scale where zero magnitude 
at 24$\micron$ 
is at 7.3 Jy.
A total of 1093 objects was detected with $[24] < 12.0 $ (0.12mJy) and S/N$ > 5$. 

\subsection {Optical Spectroscopy}

For the three possible M47 members with largest $K-[24]$ excesses 
(P1121, P369 and P922, Sec \ref{lt}), optical spectra were obtained for
us by J. Liebert 
in February 2004 using the blue channel of the 6.5-m Multiple Mirror Telescope 
spectrograph. 
With the 1$\arcsec$ slit, the resolution was 3.6 \AA$ $ and 
the wavelength coverage was 3640 -- 6790 \AA.
Exposure times were 4, 4 and 5 min for 
P1121 (V=12.7), P922 (V=14.5) and P369 (V=16.2) respectively, 
at airmass 1.8.

\section{Cluster Membership and Spitzer Detections}\label{memb}

To interpret our results, we need to determine which 24$\mu$m sources are members of the cluster. 
Our first test for membership is the color-magnitude diagram. 
To construct such a diagram, we need to
determine the cluster distance. A distance of 497 $^{+135}_{-88}$ pc
is indicated by four B stars with Hipparcos 
parallaxes \citep{Robichon1999}.
\citet{Prisinzano2003} showed that the $V$ vs. $V-I$
diagram is consistent with this distance and an age of 100 Myr.  
Estimates from Stromgren photometry include:
\citet{Shobbrook1984}: $400 \pm 10$ pc (28 B-A stars);
\citet{Nissen1988}: 425 pc (11 F stars);
\citet{RojoArellano1997}: $470 \pm 5$ pc (average) and $400 \pm 100$ pc (median) 
(36 B-F stars). 
We adopt a distance of 450 pc and $E(B-V) = 0.07$. Because the Prisinzano et al. list of 
cluster members is incomplete above 3.5M$_\odot$, we added 11 bright stars near the cluster turn-off, 
taken from \citet{Robichon1999} and \citet{RojoArellano1997}. 
Our photometrically selected sample of M47 members detected at 24$\mu$m then consists 
of 66 objects (Fig. \ref{fig1}).

We confirmed probable membership for 63 of these 66 stars by proper motion analysis.
The UCAC2 \citep{ucac2}, ACT \citep{act}, ACC \citep{allsky}, and Tycho \citep{tyc} catalogs 
provided data for all the possible cluster members except P369 (which is relatively faint). 
Only P1218 is ruled out as a probable member, but P1078 has large errors in its
proper motion and its status is unclear. All the remaining stars have appropriate
proper motions for cluster membership:
$<\mu_\alpha cos\delta>=-7.5 \pm 3.4$ mas/yr, $<\mu_\delta>=1.9 \pm 2.3$ mas/yr. 
 
\begin{figure}  
\plotone{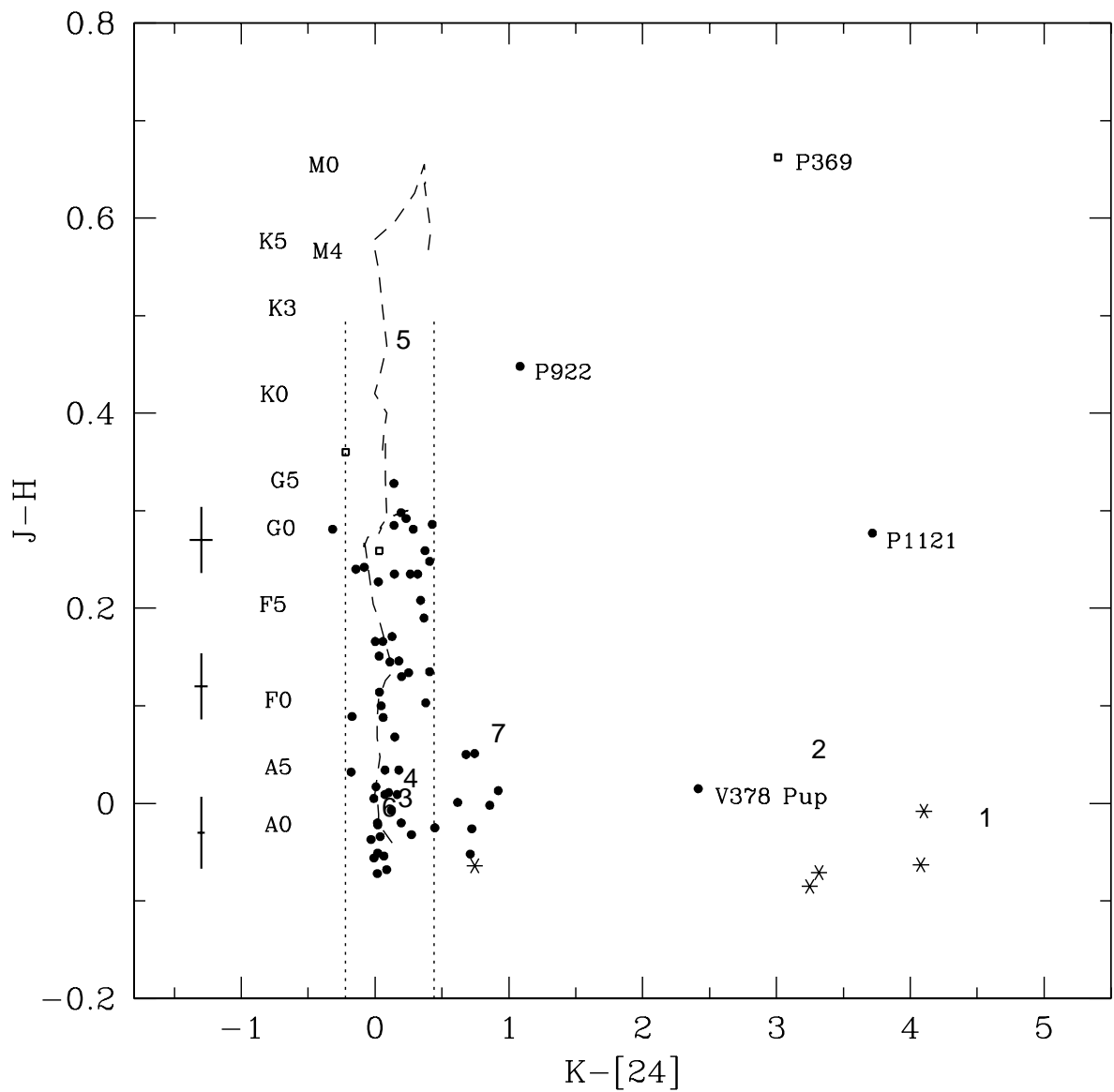}  
\caption{IR color-color diagram of M47 in comparison with Pleiades and field debris disk stars. Solid circles: M47 members; squares:
membership uncertain due to lack of proper motion information; asterisks: Pleiades stars detected by IRAS; 1: HR 4796A (A0V 10Myr), 2: $\beta$ Pic (A3V 12Myr), 3: Fomalhaut (A3V 220Myr), 4: $\beta$ Leo (A3V 240Myr), 5: $\epsilon$ Eri (K2V 330Myr), 6: Vega (A0V 350Myr), 7: $\zeta$ Lep (A2nn 370Myr). Dashed line: locus of main sequence stars converted from $K-[12]$ colors in \citet{kh95} by assuming $[24]=[12]+0.08$, where 0.08 arises because of IRAS color corrections. Dotted lines: adopted 3$\sigma$ boundaries on non-excess stars. The crosses on the left represent
typical errorbars on M47 photometry as a function of J--H (, or brightness for MS stars). 
Sources of photometry: JHK -- 2MASS and \citet{Ducati2002},
[24] -- MIPS for M47 and IRAS $[25]$ for the rest stars.}\label{fig2}
\end{figure}

We detected at 24$\micron$ all M47 members lying within the boundaries
of our mosaic and brighter than $V=12$ mag (Spectral Type $\sim$F5, 1.5M$_\odot$) 
and a few as late as $\sim$K. Accurate astrometry for the cluster members is
available both from \citet{Prisinzano2003}, who quote errors of $0.24''$,
and 2MASS, with typical errors $<$ $0.1''$. At 24$\mu$m, {\it Spitzer} 
positions are accurate to $\sim$ $1''$, but the dominant term is
an offset in the scan direction that remains fixed for a campaign.
It we take this offset out, the positional agreement between
cluster members and 24$\mu$m counterparts is better than $0.5''$.
The total probability of even one chance coincidence between random 24$\mu$m
source and any cluster member whose photospheric flux is above our detection limit (0.12mJy)
is about $P = \frac{\pi(0.5'')^{2}}{20' \times 60' \times (60 '')^{2}} \times 1093\times 63 = 1.3\%$. For objects
brighter than 1mJy at 24$\mu$m, the probability of
such a coincidence is $<$ 0.25\% (take 195 instead of 1093). Thus, it is unlikely that there
are any chance identifications for the brighter sources.   
Fig. \ref{fig2} shows $J-H$ vs. $K-[24]$ for these stars.

\section{Discussion}

The histogram of $K-[24]$ colors for the majority of M47 stars can be described by a Gaussian 
centered at $K-[24] = 0.11$ mag and with FWHM 0.26 mag ($\sigma = 0.11$ mag). 
At these long wavelengths, for pure photospheric emission one
expects $K-[24] = 0$ virtually independent of spectral type. 
Our distribution is shifted by $+$ 0.11 mag relative to this value;
however, the offset is within the systematic error of our photometry
($\sim$ 0.10 mag, arising from uncertainties in absolute calibration 
(up to 10\%) and in the aperture correction),
and the scatter is of the order of the random error of our photometry 
(0.03-0.08 mag). We therefore define excess stars as those lying redwards
of $(K-[24])_{excess} = 0.11 + 3\sigma = 0.44$ mag.
The majority of M47 stars (55 out of 66) have 24$\micron$ fluxes consistent 
with a pure photospheric origin or at most a small excess.
The remaining stars form two groups: early-type stars
with 0.6 $<$ $K-[24]$ $<$ 2.5 and three late-type stars with extreme excesses
$K-[24]$ $>$ 1. 

\subsection{Early type stars}

 Table \ref{table1} lists parameters of 8 M47 hot stars with 24$\micron$ excess 
(out of 33 total having $J-H < 0.1$, which is characteristic of A-B stars ). 
The most extreme one -- V378 Pup (= HIP 36981) -- is a blue straggler and emission-line star, 
the only star in our sample that also shows a NIR excess (Dougherty \& Taylor 1994). The incidence 
of substantial 24$\mu$m excesses is large:
24\% of the early-type stars have excesses $>$ 0.4 magnitudes.
All of these stars are brighter than 1mJy, so the possibility of a chance
coincidence with another source is very small.

It has been known since IRAS (\citet*{Waters87}, \citet{Cohen87}) that the fraction of stars
with infrared excesses in the field is the highest among B and M stars. This result is consistent
with the high incidence of excesses among the early-type stars in M47.
In the case of emission-line stars, the excess is explained by free-free processes
in the extended gaseous envelopes/disks \citep{Chokshi1987,Yudin2001}.
The nature of the excess in non-emission-line stars is not well established:
dust condensed after mass-loss episodes, dust from interstellar
cirrus, and dust from circumstellar disks may all contribute 
\citep{Kalas2002}. 

The Pleiades provide a natural comparison with M47. We identified probable Pleiades members
by searching for stars within 1.5 degrees of RA 03:47:00 and DEC +24:07:00 (2000) and
with proper motions within 7mas/yr of 17.9, -43.2 mas/yr respectively in RA and DEC. 
We used the IRAS FSC and PSC 
to identify infrared excesses in these stars.  
Our preliminary analysis of Pleiades \textit{Spitzer} data
shows that all the possible excess stars we found in this manner 
are surrounded by 0.$\arcmin$5-1$\arcmin$ extended halos that
contaminate the IRAS measurements.
Since the surface brightness from the cirrus 
between M47 stars is an order of magnitude less than in the Pleiades,
we doubt that such halos exist around M47 stars.
In addition, at the distance of M47 such halos would be resolved 
by the MIPS 6$\arcsec$ beam -- contrary to our observations (see Fig. \ref{fig1}).
This shows the importance of M47 as a 100 Myr benchmark in disk evolution,
since interpretation of Pleiades FIR data
may be severely complicated by cirrus emission.

\subsection{Late type stars}\label{lt}
 
 As one moves to later spectral types, the Vega phenomenon becomes very rare 
at wavelengths of 25$\mu$m and shorter. 
For example, \citet{Oudmaijer1992} list 462 stars with possible excesses in
the IRAS catalog. If we take dwarf stars later than F3 and earlier than K7,
brighter than V = 8, and at absolute Galactic latitude $>$ 10$\degr $ (to avoid confusion
with Galactic emission), then there are no stars with candidate excesses
greater than 1.5 magnitudes at 25$\mu$m. \citet{Mannings98} report a similar
study, with similar results. These studies searched to the full depth of
the SAO star catalog, making chance associations a possibility. As a
more stringent test, we took a complete sample
of F and G dwarfs selected to have V magnitudes that would
predict photospheric fluxes of 0.5 Jy or brighter at 25$\mu$m (and thus likely
to be measured well by IRAS). This limit was set at V = 4 for F0 to F4, V = 4.5 for F5 to F8, 
and V = 5 for F9 to G8. For these stars, we searched the IRAS PSC for high quality measurements,
and if no detection was reported there, checked if the star was listed in the FSC. Of a
total of 59 stars with good IRAS measurements, none had an excess greater than
0.3 magnitudes in [12] - [25]. Similarly, \citet{backman93} found only two F or G dwarfs
in the Bright Star Catalog (which goes to $V \sim$ 6.5) with moderate ($\sim$ 0.5 mag) 
25$\mu$m excesses (HR 506 and HR 818), and most recently, \citet{Laureijs2002}
found one K dwarf (HD191408) with excess of 0.5 mag in their 25pc volume limited survey. 

\begin{figure}  
\plotone{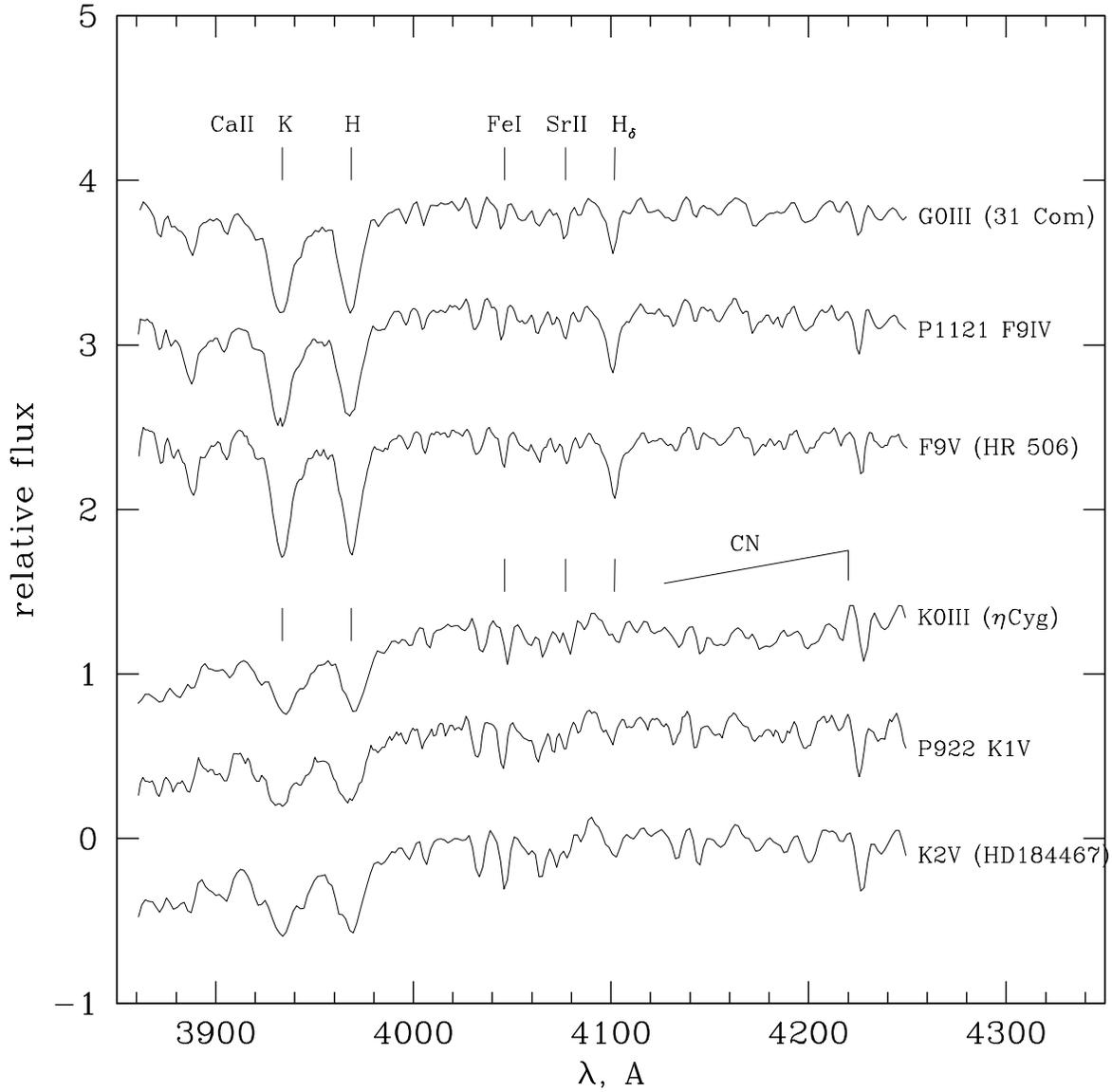}  
\caption{Portions of MMT spectra of P1121 and P922 in comparison with spectral standards 
(taken from 
NStars database). Sr II, as well as CN depression, is stronger compared to the neighbouring 
Fe I line in giants and weaker in dwarfs. Weak 
emission in the cores of Ca II lines is seen in P1121 and P922 confirming a young 
age for these stars.}\label{fig3}
\end{figure}

Therefore our discovery of three late type  M47 stars with
excesses $K-[24] = 1-4$ mag -- P1121, P922, P369 -- is very surprising.
To clarify their nature, we obtained the optical spectra of these stars shown in Fig. \ref{fig3}. 
We used spectral standards from the NStars database\footnote{http://stellar.phys.appstate.edu} to 
classify our spectra.
P369 turned out to be a reddened (A$_V \approx 1.0$ mag) K0 III-IV slightly metal poor star,
with radial velocity $\sim+100$ km/s.  It is also the only member candidate without a 
confirming proper motion measurement. Thus, we believe it is a background 
giant. On the other hand, P922 and P1121 have radial velocities consistent with
cluster membership. Their spectral types are K1.5 V and F9 IV-V respectively,
as expected from their positions on the cluster color-magnitude diagram.
Both show signs of moderate  chromospheric activity 
in CaII K and H lines (level ''k'' on scale of \citet{Gray2003}), which is a bit higher than observed in 
800 Myr old Hyades stars (R. O. Gray 2004, private communication). No emission is seen in H$\alpha$ 
but slight infilling is present in H$\beta$ for P1121.
In conclusion, P922 and P1121 have spectral signatures of young main
sequence stars consistent with being M47 members (in addition to their
proper motions and photometric characteristics).

P922 is close to the 24$\mu$m detection limit and its photosphere would
be below that limit, so the probability of
a chance coincidence of an unrelated source with some cluster member 
to produce a similar "system" is $\sim$ 8\%. Although P922 is only
slightly below the photospheric detection limit, in this calculation
we have included all known cluster members within the {\it Spitzer} image ($\sim$380). 
Therefore, a chance
coincidence is possible but not highly likely. 
P922 possesses an excess at 24$\micron$ similar to the extreme 
debris disk stars such as $\zeta$ Lep (A2 Vann, 370 Myr).

P1121 is an exceptionally interesting object. All the indicators
(photometry, spectral type, proper motion, radial velocity) place
it within the cluster. Its image (Fig.\ref{fig1}) is point-like with
no indication of extended emission in its vicinity. 
It is 7mJy at 24$\mu$m and it is above our photospheric detection
limit, so the probability of 
any chance association of a source this bright (25 out of 1093 total) with a cluster member
is only $\sim$0.03\%. 
If it is bright at longer wavelengths, it might have a disk
as massive as seen in $\beta$ Pic !
It is also possible that it is surrounded by a less-massive disk lying 
much closer to the star than is typical. At the age of P1121, 
Poynting-Robertson drag should have cleared
out most of a thick, primordial disk. It is more likely the
disk has been renewed by planetesimal collisions. This possibility is
particularly interesting given the 
arguments that the solar system went through a phase of disk
clearing, planet building, and violent planetesimal collisons
for its first few hundred Myr \citep{Hartmann2000}.

\section{Conclusions}

We find substantial infrared excesses at 24$\mu$m in 10 stars that are members of 
the 100 Myr old cluster M47. The most interesting example is P1121, which is of
spectral type F9 IV-V, yet has $K - [24]$ $\sim$ 3.7. If this excess is due to
a debris disk, it likely has a large mass and is a very rare object judging from
IRAS and ISO observations of the debris-disk phenomenon.

\acknowledgments

We are grateful to Jim Liebert and Curtis Williams for obtaining spectra
of the IR-excess stars, Richard Gray and Chris Corbally for help with spectral classification,
and Eric Mamajek for useful discussions.
This work is based on observations made with the Spitzer Observatory, 
which is operated by the Jet Propulsion Laboratory, California Institute of Technology 
under NASA contract 1407. Support for this work was provided by NASA through 
Contract Number 960785 issued by JPL/Caltech.

\clearpage  
\begin{deluxetable}{lllrrrrlllrrr}
\tablecolumns{7} 
\tablewidth{0pt}
\tabletypesize{\small}
\tablecaption{24{\micron} excess stars in M47\label{table1}}
\tablehead{ 
\colhead{P.N} & \colhead{Name} & \colhead {SpT} & \colhead {M$_{V}$} & 
\colhead {J-H} & \colhead {K-[24]} & \colhead {F$_{24}$ (mJy)} }
\startdata
\\
\nodata & HIP 36981  & B2-B5 Ven &   -2.61 &  0.02  & 2.42 & 558.89 $\pm$ 0.51 \\
\nodata & HIP 36967  & B5-B9 Vn  &   -0.33 & -0.05  & 0.71 & 8.86 $\pm$ 0.05  \\
\nodata & HD 60995  & B8-B9 V   &    0.38 & -0.03  & 0.72 & 5.19 $\pm$ 0.05 \\ 
 1152 & 2MASS 07371449-1425179 &  \nodata    &    1.73 &  0.00  & 0.62 & 1.58 $\pm$ 0.03 \\
 1155 & BD -1402028  & A2p      &    2.07 &  0.01  & 0.92 & 1.67 $\pm$ 0.02 \\
1173 & 2MASS 07362205-1430283 &  \nodata    &    2.13 & -0.00  & 0.86 & 1.55 $\pm$ 0.02 \\ 
1172 & 2MASS 07365524-1433176 & \nodata    &    2.42 &  0.05  & 0.68 & 1.13 $\pm$ 0.03\\
 1182 & 2MASS 07370687-1431348 &  \nodata    &    2.75 &  0.05  & 0.74 & 1.14 $\pm$ 0.03\\
 \\
 1121 & 2MASS 07354269-1450422 & F9 IV/V  &    4.40 &  0.28  & 3.72 & 7.31 $\pm$ 0.03\\
  922 & 2MASS 07364576-1434348 & K1 V     &    6.27 &  0.45  & 1.08 & 0.23 $\pm$ 0.02 \\ 
 369\tablenotemark{a} &    2MASS 07372200-1424262          & K0 III/IV & \nodata &  0.66  & 3.01 & 0.94 $\pm$ 0.03 \\ 
\\
\enddata
\tablenotetext{a}{Our spectrum indicates it is not a cluster member.}
\tablecomments{P.N -- numbering sequence from Table 4 of \citet{Prisinzano2003}; SpT -- from the literature \citep[VizieR;][]{WEBDA,Hartoog,Dworetsky}, except for late types (this work); M$_{V}$ -- assuming distance 450pc}
\end{deluxetable} 

\clearpage

\clearpage  
 
\clearpage  

\end{document}